\documentclass{singlecolPressArxiv}
\usepackage{mathrsfs}
\usepackage{graphicx}
\usepackage{amsmath,upgreek}
\usepackage[square,sort,comma,numbers]{natbib}
\makeatletter
\renewcommand\@biblabel[1]{#1.}
\makeatother

\theoremstyle{TH}{

}

\theoremstyle{THrm}{

}

\theoremstyle{THhit}{

}

\makeatletter
\def\theequation{\arabic{equation}}

\makeatother

\begin{document}%
%%%%%%%%%%%%%%%%%

\setcounter{page}{1}

\LRH{E. Prati}

\RRH{Atomic scale nanoelectronics for quantum neuromorphic devices}

\VOL{x}

\ISSUE{x}

\PUBYEAR{2016}

\BottomCatch

%\CLline

\PUBYEAR{2016}

\subtitle{\textit{Author's Original}}

\title{Atomic scale nanoelectronics for quantum neuromorphic devices: comparing different materials}

\authorA{Enrico Prati}
\affA{Istituto di Fotonica e Nanotecnologie, Consiglio Nazionale delle Ricerche\\ Piazza Leonardo da Vinci 32, I--20133 Milano, Italy\\
Fax:  +39 02 23996126 \qquad E-mail: enrico.prati@cnr.it}
\begin{abstract}
I review the advancements of atomic scale nanoelectronics towards quantum neuromorphics. First, I summarize the key properties of elementary combinations of few neurons, namely long-- and short--term plasticity, spike-timing dependent plasticity (associative plasticity), quantumness and stochastic effects, and their potential computational employment. Next, I review several atomic scale device technologies developed to control electron transport at the atomic level, including single atom implantation for atomic arrays and CMOS quantum dots, single atom memories, Ag$_2$S and Cu$_2$S atomic switches, hafnium based RRAMs, organic material based transistors, Ge$_2$Sb$_2$Te$_5$ synapses. Each material/method proved successful in achieving some of the properties observed in real neurons. I compare the different methods towards the creation of a new generation of naturally inspired and biophysically meaningful artificial neurons, in order to replace the rigid CMOS based neuromorphic hardware. The most challenging aspect to address appears to obtain both the stochastic/quantum behavior and the associative plasticity, which are currently observed only below and above 20 nm length scale respectively, by employing the same material.  
\end{abstract}

\KEYWORD{Artificial Neurons; Artificial Synapses, Atomic Scale Nanoelectronics; Quantum Neuromorphics; Plasticity.}

\begin{bio}
Enrico Prati received his PhD in Physics at
the University of Pisa in 2002. He is Research Scientist of Istituto di Fotonica e Nanotecnologie
of Consiglio Nazionale delle Ricerche in Italy, and Visiting Scholar of Waseda Univesity Tokyo.
His research interests include applications of atomic scale nanoelectronics
to neuromorphic and photonic systems. He is author of a number of research studies published on high impact factor journals, as well as book chapters. He has been Co-Editor of a book on atom based nanoelectronics. He has been Keynote Speaker of IEDM and panelist of ITRS on deterministic doping of nanoelectronic devices.
\vs{9}
\end{bio}

\maketitle
\section{Introduction}

Neuromorphic devices have been inspired by the increasing understanding of mechanisms underlying the behaviour of real neurons, but hardware architectures have been principally focused on a brute force method to obtain realistic transfer functions of solicited artificial neurons (see \cite{Izhikevitch04}) and neural networks. Emulation of individual natural ingredients by using artificial components \cite{Lizeth12} has been generally neglected until the recent inclusion in the Emerging Research Device Chapter of ITRS 2013 \cite{ITRS}.
Both classical and quantum schemes have been proposed to achieve artificial neuron behavior \cite{Indiveri11} and quantum adiabatic neural systems \cite{Kinjo05} irrespectively from their biological plausibility. On the other hand, there are realistic (naturally inspired) kinds of neurons like those of \cite{Maeda00} which can be developed from both the material science and the architectural point of view. The implementation of biologically plausible artificial neurons and synapses at nanometric scale requires the development of techniques based on quantum chemistry, quantum mechanics, atomic physics and material science. The methods for emulating specific properties of neurons are based on radically different approaches and, to date, the topic has never been systematically reviewed. In this review paper we separately consider the key properties of a biological neural system including neurons and synaptic activity, and the implementation realized by exploiting different materials. 
\\
Real neurons and synapses are based on molecular scale phenomena, including ionic transport, single charge fluctuations, protein folding, as well as genic expression of DNA. At such nanometric scale, quantum mechanics plays a major role. Studies in microtubules seems to confirm such conjecture \cite{Sahu13}. Even if quantum effects have not been definitively  demonstrated in neurons and in human brain in general, recent reports of quantum effects at room temperature in biological systems suggest that similar effects may be spread in many of them \cite{Nori13}, including the brain, and the field is growing rapidly. 
In this paper, the recent progress in the fabrication of atomic scale devices suitable for implementation of nanometric electric functionality of real neurons is discussed.
\\
Artificial neurons can be described in terms of blocks which emulate functional properties. A nature-inspired artificial neuron will provide convergence and integration among the different blocks in a possibly single platform (namely materials and process). The blocks include synapse blocks (responsible of integration of the input spikes, elaboration of temporal dynamics, short and long-term plasticity mechanisms and of providing an output signal), a soma block (for temporal integration, spike generation, refractory period block and spike-frequency or spiking threshold adaptation block), and axon and dendrites spatial structure block (to implement the cable equation, for modeling signal propagation along passive neuronal fibers).
Generally the literature is concentrated to synaptic emulation as plasticity is considered of greater complexity if compared to aspects related to connectivity and internal mechanisms of signal processing in neurons. The latter can 
in principle be replaced by an equivalent circuit with similar transfer function and connections.
In the first part of this review, the key properties to achieve a nature inspired neural system are considered. In the second part a number of different fabrication methods and materials are reviewed. The features achieved by the fabrication methods are compared in the Discussion section. A connection emerges between fluctuations and quantum effects with nanometric scale, and between associative plasticity and size above 20 nm.

\section{Requirements for nature inspired artificial synapses} \label{sec:natureinspired}
This section is devoted to the different kinds of plasticity and stochastic processes existing in real neurons and synapses. 
The classification here proposed is based on the number of neurons involved in the process, and the time evolution of both the voltage solicitations and the consequent changes of the conductance. The timescale of the duration is also considered. In the following the term plasticity is used to refer to synaptic plasticity, not to be confused with intrinsic plasticity. The latter, not discussed in this work, consists of a persistent modification of the electrical properties of neurons, as a consequence of neuronal or synaptic activity. Long term plasticity, short term plasticity, spike-timing dependent plasticity, quantumness and stochastic activation are considered. Their computational uses are summarized in Table 1.

\subsection{Long term plasticity}
Long term plasticity involves either two or three neurons, including a presynaptic neuron A1 (eventually a modulatory interneuron neuron A2) and a postsynaptic neuron B. 
Long term plasticity may last either hours or days.  The induction of long term plasticity results in either an increased or a decreased probability of discharge of postsynaptic neurons after repeated potential bursts of presynaptic neurons. 
\begin{itemize}
	\item \textbf{Long term potentiation}
Long term potentiation (LTP) involves two neurons, a presynaptic neuron A1 which acts as a generator of action potentials, and a postsynaptic neuron B. It results in the increase of the synaptic strength when activated by a high frequency stimulus. 
	\item \textbf{Homosynaptic long term depression}
Homosynaptic long-term depression (LTD) involves two neurons, a presynaptic neuron A1 which acts as a generator of action potentials, and a postsynaptic neuron B. It results in the weakening of the synapse when activated by a low frequency stimulus. 
  \item \textbf{Heterosynaptic long term depression}
Differently from the previous homosynaptic LTD, heterosynaptic LTD involves three neurons. It occurs at synapses that are not potentiated or are inactive between A1 and B. The weakening of the synapse is controlled by the firing of a distinct modulatory interneuron A2, independently from the activity of the presynaptic or postsynaptic neurons, as reported by \cite{Escobar07}
\end{itemize}
In the literature on artificial neurons, the term long-term plasticity is often improperly used to refer to long term potentiation. Usually long term potentiation and depression are a byproduct of the spiking time dependent plasticity (STDP) scheme (section 2.3), where three neurons are involved, two of which fire. From a computational point of view, LTP is employed by nature to encode space \cite{LTP}, while LTD to encode the features of space \cite{LTD4}, for example orientation and finer details \cite{LTD3}. In addition, LTD may determine selective weakening of synapses \cite{Purves} and the clearing of old memory traces (see \cite{LTD1} and \cite{LTD2}.

\subsection{Short term plasticity}
Unlike with long term plasticity, short term plasticity (STP) involves only two neurons, namely a presynaptic neuron A, which has the role of exciting a postsynaptic neuron B. STP invokes a change in the electrical properties of a single synapse, and it can act both in a depressive and in a facilitative mode. 
The modification to synaptic efficacy is temporary and lasts generally 100-1000 ms. Both the facilitation and the depression depend on the local finite amount of ions and molecules at the synapse, which are continuously exchanged and re-employed at every process in the site. It is worth noting that the convention here employs P for \textsl{plasticity}, while in LTP it refers to \textsl{potentiation}. For historical reasons, the analogous of the potentiation is called \textsl{facilitation }for the short term processes.
\begin{itemize}
	\item \textbf{Short term facilitation}
Short--term facilitation (STF) constitutes an increased excitability of a synapse as reaction to an external stimulus. STF is due to penetration of Ca ions in the axon terminal after having generated a spike, with a consequent increase in the release of neurotransmitters. 
 
	\item \textbf{Short term depression}
Short--term depression (STD) consists in a diminished reaction of a neuron to external stimuli, due to the depletion of neurotransmitters during the synaptic process at the axon terminal of a pre-synaptic neuron. 
\end{itemize}

STF-enhanced synapses can be exploited to hold the memory trace of an input without recruiting persistent firing of neurons, which makes for a robust method to implement working memory \cite{STF1}. \cite{STF2} showed that STF can map the input features from a low-dimensional space to the high-dimensional state space of a network to boost read-out of input information. STD can be employed to remove auto-correlation in temporal inputs, by exploiting the depression effect to reduce the output correlation of post-synaptic potential \cite{STD}.

\subsection{Spike timing dependent plasticity}\label{STDP}
Spike Timing Dependent Plasticity (STDP) is a temporally dependent asymmetric form of plasticity, which realizes associative plasticity. 
It involves at least three neurons, as the simplest system is constituted by a presynaptic neuron A1 coupled with postsynaptic neuron B, whose conductivity is changed after the process is completed, and a control presynaptic neuron A2 that excites the firing of the neuron B at some time. STDP is found in the following contexts:
\begin{itemize}
	\item \textbf{Forward STDP} (F-STDP) An asymmetric change in the conductance where the arrival of a repeated presynaptic spike from A1 a few milliseconds before postsynaptic action potentials in B leads to LTP, while the repeated arrival of presynaptic spikes after the occurrence of postsynaptic spikes in B leads to LTD. The connection between A2 and B does not change in the process. 
 \item \textbf{Reversed  STDP} (R-STDP) Similar to F-STDP, where the repeated presynaptic spike arrival of a repeated presynaptic spike from A1 a few milliseconds before postsynaptic action potentials in B leads to LTD, and repeated spike arrival after postsynaptic spikes in B leads to LTP, like in some inhibitory connections to neocortical pyramidal neurons and corticostriatal synapses. 
 \item \textbf{Invertible STDP} (I-STDP) In some cases, STDP has a major dependence on the voltage of the postsynaptic neuron B just before the generation of action potentials. The voltage at the time of the action potential is able to control the direction of the change of the synapse, even for fixed spike timing \cite{Sjostrom01}. Depending on the voltage in B, the STDP behaves as either forward or reversed STDP.
\end{itemize}
STDP is often used to refer to F-STDP. F-STDP has been achieved in several systems (see the Discussion section). Both forward and reversed STDP have major computational use. F-STDP is considered the main faster responsible for memorization. F-STDP includes possible employment based on the emergence of long-range temporal correlations \cite{Corcoran} and temporal coding \cite{STDP3} or spacetime localization learning (see \cite{STDP1} and \cite{STDP2}).  R-STDP produces anti-Hebbian synaptic plasticity which allows to predict and cancel self-generated sensory signals (see \cite{ISTDP2} and \cite{ISTDP1}).  

\begin{table}[th]
\center
\small
\caption{Various properties of synaptic activity are used in the brain for computational purposes. The table lists the key properties, the possible computational use and the references.}
\label{Table1}
\begin{tabular}{lllcc}
\hline
Property&Computational use&References\\
\hline
STF&Working memory & \cite{STF1} \\
   &Mapping input for readout & \cite{STF2} \\
STD&Removing auto-correlation& \cite{STD} \\
LTP&Spatial memory storage& \cite{LTP}, \cite{LTD3}\\
LTD&Encoding space features& \cite{LTD3}, \cite{LTD4} \\
   &Selective weakening of synapses& \cite{Purves}\\
	 &Clearing old memory traces& \cite{LTD1}, \cite{LTD2}\\
F-STDP&Long range temporal correlation&\cite{Corcoran}\\
      &Temporal coding&\cite{STDP3}\\
      &Spatiotemporal coding&\cite{STDP1}, \cite{STDP2}\\
R-STDP&Sensory filtering&\cite{ISTDP1}, \cite{ISTDP2}\\
SA/Q&Enhanced excitability&\cite{Sobie11}\\
\hline
\end{tabular}
\end{table}

\subsection{Stochastic activation / Quantumness}
The response of neurons to external stimuli is subject to some degree of randomness due to electrical spatially distributed noise. Channel noise, generated by the random gating of voltage-gated ion channels, has measurable effects on the dynamics of single neurons and it increases the range of spiking behaviors exhibited in a neural population \cite{White00}. Current fluctuations of individual ions and electrons manifest themselves at the level of a single quantum of conductance G$_0$. Such effects may involve a number of neurons ranging from one to many. In addition, interference from noise of larger amplitudes with a spike train occurs even when the noise and signal do not overlap in time. Weak noise in neurons may inhibit the generation of bursts of spikes but only so when they overlap in time (see \cite{Tuckwell10}). We refer to such a broad class of phenomena with stochastic activation/quantumness (SA/Q) in Tables 1 and 2. \cite{Maass} has shown how noise provides a powerful resource to enhance computational capability in neuromorphic systems.
Nature exhibits underlying noise assisted activation, as suggested by \textit{in vivo }experiments \cite{invivo}. 
According to \cite{Sobie11}, neuron response times are optimal under static noise that is sharply peaked at zero frequency, as a large number of neurons are close to threshold just before a step stimulus arrives.

\section{Hardware implementation of atomic scale neuromorphic properties} \label{sec:hardware}
Several fabrication methods have been proposed to implement short--term and long--term plasticity and spike--timing dependent plasticity. This section reviews the different methods, including Ge$_2$Sb$_2$Te$_5$ (GST) phase change materials, silicon, aluminum single atom memories, Ag$_2$S and Cu$_2$S atomic switches, hafnium based RRAMs, organic material based transistors, to implement such features, as well as other features related to stochastic activation of synapses and quantum effects. 

\begin{figure*}[t]%2
\caption{F-STDP (left, black) and R-STDP (right, red) in GST from Ref. \cite{Kuzum11}. In the convention adopted here, $\Delta t >0$ when the presynaptic spike occurs before the postsynaptic spike. In F-STDP, the synaptic weight is increased when the presynaptic spike comes first (left). Reprinted with permission from Kuzum, D., Jeyasingh, R. G., Lee, B., and Wong, H. S. P. (2011), 'Nanoelectronic programmable synapses based on phase change materials for brain-inspired computing', \textit{Nano letters}, Vol. 12, No. 5, 2179--2186. Copyright (2012) American Chemical Society.}\label{fig:STDP}
\centerline{\includegraphics[scale=0.33]{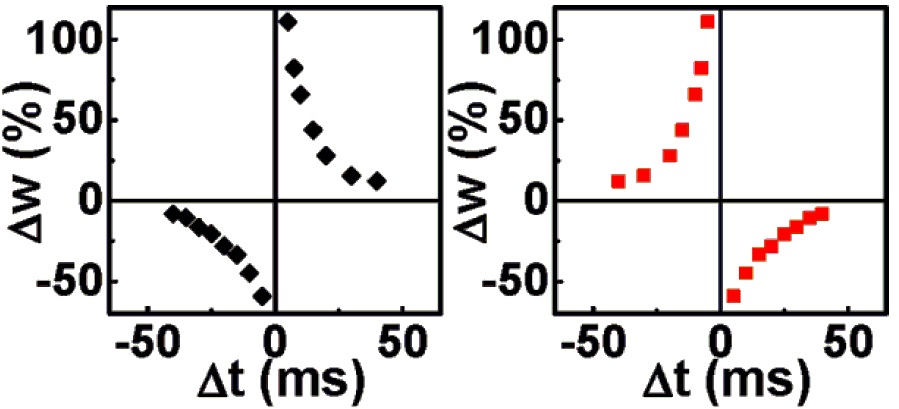}}
\end{figure*}

\subsection{Nanoelectronic programmable synapses based on phase change materials}
Synapses based on phase change materials (PCM) reported by \cite{Kuzum11} have been developed as an
alternative to CMOS based architectures. It operates at pJ energy scale, while CMOS requires a  
larger power consumption (at least 10 transistors per synapse would be required for an equivalent response). PCM are sufficiently compact to overcome scalability constraints and $10^{15}$ circuits can be targeted. 
The operation principle relies on the switching between amorphous (reset, high resistivity) 
and crystalline (set, low resistivity) states by applying electric pulses capable of locally heating 
the device and inducing a phase transformation.
GST has been used by \cite{Kuzum11} to create artificial synapses. 
The device consists of GST deposited between a 75 nm diameter bottom electrode and a top
electrode (20 nm is the minimum feature size of this approach).
PCM are ordinarily programmed to up to 16 intermediate resistance levels
by current pulses. Differently, for synaptic applications, a finer resistance control (1 \% change) is performed. An order of magnitude change in the
phase change cell resistance has been reported through 100 steps for
both the set and reset transitions. Both F-STDP and R-STDP have been demonstrated, as shown in Figure 1.

\subsection{CMOS and silicon nanoelectronics}
Currently, CMOS technology provides the only platform which includes the minimum set of properties to emulate an individual neuron and a network of neurons. Silicon neurons can be described as circuits having one/multiple synapse blocks, receiving spikes from other neurons (time--integrated and converted in currents), a soma block (time--integrating inputs and generating the output action potentials or alternatively digital spikes), and circuits emulating the network of dendridic trees and axons.
The circuits process short-- and long--term plasticity mechanisms, conversion of voltage spikes into excitatory and inhibitory post--synaptic currents (EPSCs and IPSCs), soma integration and some adaptation functionalities. 
Such functionalities generally rely on the robust architecture of the 180 nm CMOS node, where device variability does not affect the performances. Inspired by the development of the 28 nm technology node of IBM TrueNorth silicon chip \cite{Modha}, we investigate nanoscale effects offered by the technology to directly implement complex properties of neurons and synapses in the devices, instead of using large circuits built with many components.
Few atom \cite{Mazzeo12,Prati12} and few--electron devices \cite{Prati12b} at the sub 14 nm node induce both strong non--linear effects, variability and stochastic fluctuations at room temperature. Random fluctuations generated by charged defects \cite{Prati07,Prati08} may generate Lorentzian spectrum noise which provides an internal source of neuron activity and neural synchronization \cite{Lizeth12}. At either cryogenic temperature or size below 5 nm at room temperature, transport is dominated by quantum regime. Single electron circuits for depressing synapse have been explored to develop noise--tolerant architectures \cite{Oya06}. Native memory and plasticity effects are not reported in silicon technology.
\begin{figure*}
\caption{STF (left) and LTP (right) of Ag$_2$S atomic switch, in \cite{Ohno11}. Reprinted with permission from Ohno, T., Hasegawa, T., Tsuruoka, T., Terabe, K., Gimzewski, J. K., and Aono, M. (2011) 'Short-term plasticity and long-term potentiation mimicked in single inorganic synapses', \textit{Nature Materials}, Vol. 10, No. 8, pp. 591--595. Copyright (2011) Nature Publishing Group.}\label{fig:Figure1}
%\centerline{\epsffile{Figure1.eps}}
\centerline{\includegraphics[width=\textwidth]{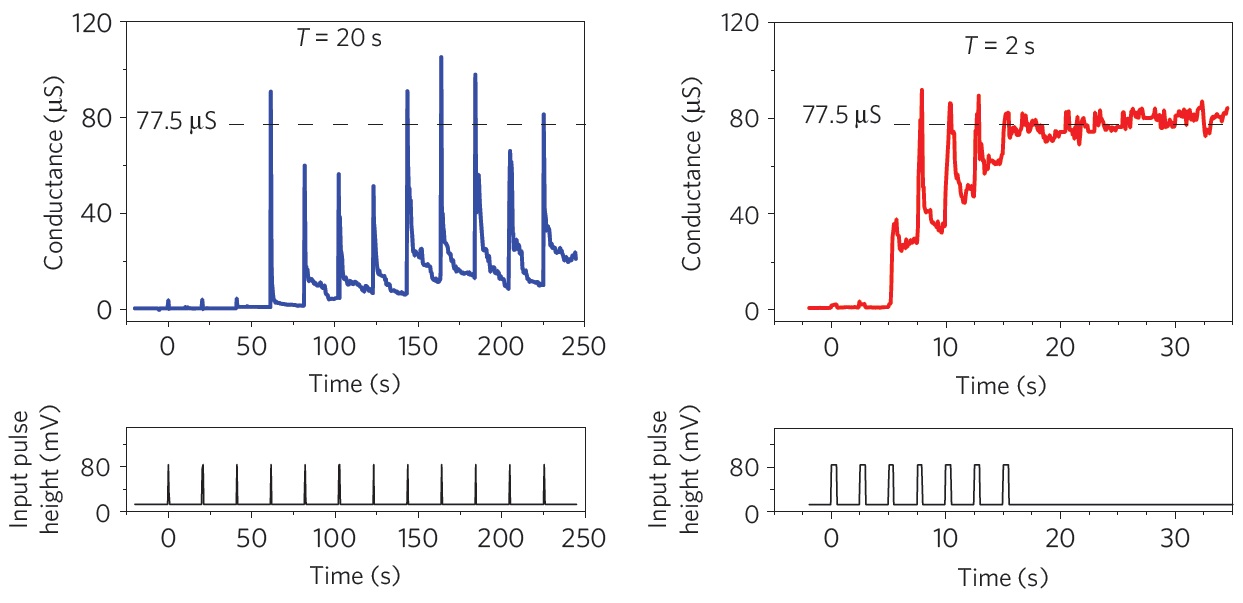}}
\end{figure*}
\subsection{Ag$_{2}$S atomic switch}
A viable way to bypass the difficulty of silicon technology issues is based on
the implementation of a different kind of plasticity, such as STF
and LTP at the physical level. The
change in conductance is considered analogous to the change in strength 
of a biological synaptic connection (synaptic weight). 
It has been reported by \cite{Ohno11} that an Ag${_2}$S inorganic synapse is able to emulate synaptic 
functions with both STF and LTP characteristics through the use of input pulses repeated in time (Figure 1).
Ag${_2}$S form an atomic switch, which operates at certain critical voltages, and stores information 
as STF with a spontaneous decay of the conductance according to intermittent input stimuli. 
Conversely, a more frequent stimulation determines a transition to LTP. The key point is that 
Ag${_2}$S inorganic synapse emulates a biological synapse to achieve
dynamic memorization in a single device, with no need of
external pre--programming. 
Such an implementation can be seen as a redox version of a memristor. 
The working principle consists of the application of the input pulses to determine the precipitation of Ag atoms from an Ag$_{2}$S electrode, to form an Ag atomic bridge between the electrode itself and a counter metal electrode. If the precipitated Ag atoms are insufficient to form a bridge, the inorganic
synapse works as STF. If they form a bridge, it works as LTP. Similarly to a biological synapse, for which the release of neurotransmitters is due to
the arrival of firing caused by action potentials, the signal is transmitted as a synaptic potential. In addition, a frequent stimulation 
enhances the strength of the synaptic connection at a long--term time scale.

\subsection{Cu$_{2}$S atomic switch}
Similarly to the Ag$_{2}$S atomic switch, it has been demonstrated that also a Cu$_{2}$S gap--type 
atomic switch is also able to emulate the synaptic short--term and long--term plasticity as reported by \cite{Nayak12}. 
As in the previous material, the plasticity here is controlled
by the interval, the amplitude, and the width of input voltage
pulses. The degree of air or moisture in the environment influences the plasticity, 
which is also demonstrated by time-dependent scanning tunneling microscopy images 
of the Cu-protrusions. A higher temperature increases the long-term memory effectiveness, 
as shorter or fewer are pulses, similarly to biological systems. 
The Cu$_{2}$S gap-type atomic switch is obtained by a Cu$_{2}$S 
solid electrolyte on a Cu electrode
and a counter Pt electrode separated by a nanogap between
them by a scanning tunneling microscope. A voltage
applied between the electrodes such that Cu$_{2}$S is at the positive
bias, makes the Cu$^{+}$ ions (initially uniformly distributed) diffuse toward the Cu$_{2}$S surface.
The temperature and the voltage control the precipitation of Cu$^{+}$ ions on the
surface, because of the electrochemical reaction Cu$^{+}$ + e$^{-}$ $\rightarrow$ Cu.
Next , the precipitated Cu atoms form a bridge between
the electrodes, which increases the conductance. Each single atomic contact saturates the quantum of 
conductance $G_0 = 2 e^2 / h = 77.5$ $\mu$S, where e is the electron charge and h is
Planck's constant. 
The inorganic synapse shows three different conductance states, similar to sensory memory (SM),
STF, and LTP of a real synapse. Differing from Ag$_{2}$S, which is an n-type material, Cu$_{2}$S 
is a p-type material as Cu vacancies act as electron acceptors where free holes contribute to conductivity 
(the resistance increases by increasing the Cu concentration). 

\subsection{Al single atomic switch}
\cite{Shirm12} have proposed a two--terminal (rather than three--terminal) electronic switch based on a metallic atomic-scale aluminum contact to be operated as a reliable and robust switch. Through the use of an electromigration protocol, the conductance of the aluminum atomic contact
is controlled between two values in the range of a few conductance quanta. Such change is ultimately associated with a switching process caused by the reversible rearrangement of single atoms, as revealed by current-voltage characteristics associated with superconductivity with molecular dynamics.
The behaviour is hysteretic (with two distinct states), so the two-terminal switch may be employed as a non-volatile information storage element.
When the switch is operated, the geometry changes from a single-atom contact to a
dimer contact, with the rearrangement of a single atom and the
rupture of only two bonds.  
The operation as a memory device is based on a potential read-and-write scheme: with a bistable
state, whereby the control current is raised and lowered abruptly. The two respective conductance values are identified with the 1 state and the 0 state respectively, similarly to a flip flop, and they range between 1 and 2 $G_0$.
Thanks to the hysteretic nature of the switching process, small current biases can be used for the read phase, and high negative or positive biases for the write phase. 

\subsection{InGaZnO Memristors}
Among the memristors proposed over the years based on ion migration or atomic   
switch mechanisms, such as TiO$_2$ , Ag$_2$S, Cu$_{2}$S, RbAg$_4$I$_5$ (see below), SiAg, and WO$_x$ as single synaptic devices, the InGaZnO memristor of
\cite{Wang12} emulates not only the STDP and transmission characteristics of the synapse, 
but also includes spike-rate dependent, long-term/short-term plasticity and learning-experience behavior. 
The learning-experience function is related to the metastable local structures
in amorphous $\alpha$-IGZO. A frequent stimulation may 
cause an enhancement of LTP, both spike-rate-dependent and spike-timing dependent
plasticity, and the STF to LTP transition may occur from repeated stimulation. 
The physical mechanism is based on ionic currents from oxygen migration/diffusion. Both LTF and STP are of the same time scale as in the human brain. The learning process is observed by the re-stimulation process from a mid-state. Far fewer pulses than the number of stimuli required in the first learning process are needed by
the device for recovering its memory, like in the learning process based on experience in biological systems.

\subsection{HfO$_2$-based Resistive switching memory}
Resistive switching memory (RRAM) devices are based on the change of 
the resistance when a metal is oxidized in response to 
electrical pulses. The physical process is based on a 
set and a reset transition in an oxide RRAM. 
The initial state for the set transition is the reset state, 
where a conductive filament (CF) is interrupted by a gap as
a result of ion migration under negative voltage. The gap is
depleted from oxygen vacancies and metallic impurities, such as excess
Hf, responsible for the enhanced local conduction in the CF.
The defects are generated by the dielectric breakdown 
needed to initiate resistive switching at the formation process. 
The RRAM reset state has a high resistance and when a positive voltage 
is applied to the top electrode the ionized defects migrate from the CF region close to the top electrode, with a resulting increased conductance, like a growth of the diameter of the connecting
sub-filament within the gap.
It is worth to remark that the method above radically differs from those merely based on a resistive switching device, integrated in a larger circuit including an LTP block, an LTD block, and a peripheral circuit, as the LTP and LTD effects are natively carried by the device.
In addition, random telegraph signals have reportedly been used in RRAMs by \cite{Puglisi12,Ambrogio15}, enabling the inclusion of stochastic activation.
To be employed in neuromorphic synapses, an RRAM 
has to fulfill two key requirements: to provide a resistive connection between a
presynaptic neuron and a postsynaptic neuron and to provide tuneability of its conductance 
in response to the neuron activity. The learning process may take place as STDP , i. e., as LTP when a postsynaptic pulse follows the presynaptic pulse, as long-term
depression (LTD) in the reversed case, where the conductance decays 
exponentially with the delay between the pre- and postsynaptic pulses. 
\cite{Ambrogio13} have demonstrated an STDP circuit consisting of
a one-transistor-one-resistor (1T1R) structure comprising a
MOSFET and a HfO$_2$-based RRAM, included in a larger circuit providing 
the presynaptic and the postsynaptic artificial neurons. 

\subsection{Organic ionic/electronic hybrid materials in synaptic transistor}
Finally, we consider organic media for synaptic-like behavior \cite{Lai10}.
Such an approach has the interesting feature that, similarly to phenomena in a real brain, 
the signals may be transmitted by means of ionic fluxes. Indeed, potential spike signals in a
presynaptic neuron can trigger an ionic excitatory postsynaptic
current (EPSC) or inhibitory postsynaptic current (IPSC).
The postsynaptic neurons collectively process such currents through 
$10^{3}$--$10^{4}$ synapses which control spatial and
temporal correlated functions, whose efficacy is modified by temporally correlated pre-- and
post--synaptic spikes via STDP. LTP can be obtained if a postsynaptic spike occurs after a presynaptic spike by a few milliseconds, which increases synaptic efficacy, whereas LTD occurs if the two spikes timings are reversed. Through the integration of a layer of an ionic conductor and a layer of ion-doped conjugated polymer onto the gate of a Si-based transistor, a synaptic transistor based on
ionic/electronic hybrid materials has been obtained. A potential
spike triggers ionic fluxes with a temporal lapse of a few
milliseconds in the polymer, which in turn spontaneously
generates EPSC in the Si layer. The ions stored in the polymer are modified by means of 
temporally correlated pre- and post-synaptic spikes, which, like for real synapses, results 
in a strengthening or weakening of the device transmission efficacy with STDP. 
An example of such a synaptic transistor is based on a standard Si n-p-n source-channel-drain
MOS, with a 3-nm-thick SiO$_2$ gate oxide. A 70-nm-thick conjugated polymer layer of poly[2-methoxy-5-
(20-ethylhexyloxy)-p-phenylene vinylene] (MEH-PPV) and a
70-nm-thick ionic conductive layer of RbAg$_4$I$_5$ are sandwiched
between the gate oxide and an Al/Ti electrode.
The presynaptic spikes are applied to
the transistor gate, and postsynaptic currents are measured
from the source while postsynaptic spikes are also applied to the
source. The typical time scale is of the order of milliseconds. 
\begin{table}[th]
\center
\small
\caption{Neural properties and device technology. On the horizontal axis, the properties required by artificial synapses: stochastic activation/quantumness, short term facilitation, short term depression, long term potentiation, both kinds of long term depression, forward STDP, reverse STDP, invertible STDP and scale size of the device in nanometers. On the vertical axis, the device technology  considered in section 2}
\label{Table2}
\begin{tabular}{lccccccccccc}
\hline
						&SA/Q&STF&STD&LTP&LTD&F-STDP&R-STDP&I-STDP&size\\
Neurons     &1--N& 2 & 2 & 2 & 2--3&3   &   3  &   3  & nm\\
\hline
Al atom		&$\bullet$ &  &  &  &  &   &  &  & 1-2\\
Ag$_{2}$S	&$\bullet$ & $\bullet$&  & $\bullet$&  &   &  &  & 1-2\\
Cu$_{2}$S	&$\bullet$ & $\bullet$&  & $\bullet$&  &   &  &  & 1-2\\
Si				&$\bullet$ &  &  &  &  &   &  &  & 2-130\\
HfO$_2$ RRAM&$\bullet$& &  &  &  &  $\bullet$&  &  & 20 \\
InGaZnO		&  & $\bullet$&  & $\bullet$&  &  $\bullet$&  &  & 20-40\\
GST 			&  &  &  &  &  &  $\bullet$& $\bullet$& $\bullet$& 20-100\\
Organic Hyb. && &  &  &  &  $\bullet$&  &  & 20-200\\
\hline
\end{tabular}
\end{table}

\section{Discussion}
The correspondence between the quantum neuromorphic devices and biological neurons leads to a table summarizing the properties which has been artificially covered. Pure LTP from the two-neurons scheme has been reported with memristors based on Ag$_2$S and Cu$_2$S, as well as in the InGaZaO memristor, while pure LTD has not been observed.  Figure \ref{fig:Figure1} (right) shows the realization of LTP in Ag$_2$S. STF has also been reported in Ag$_2$S. The same materials that allow LTP also reproduce STF. Like LTD, neither is pure STD currently reported. Figure \ref{fig:Figure1} (left) shows the realization of STF in Ag$_2$S.
\\
F-STDP has been demonstrated in InGaZnO memristors, HfO$_2$ based RRAMs, the organic hybrid method, and GST. GST is currently the most complete method for STDP, as all the F-STDP, the R-STDP and the I-STDP have been realized experimentally.
Interface and oxide defects in SiO$_2$ and HfO$_{x}$ and single atom doping in CMOS technology are currently the only method to introduce bistable noise at a native level in the components, like what happens in ion channels. Current effects at a G$_0$ conductance level are observed in silicon quantum dot technology, in the Al single atom memory, and in Ag$_{2}$S and Cu$_{2}$S atomic switches.

Table 2 shows the potential of various device technologies for implementation of neural mechanisms. The second line indicates the number of neurons involved in the process. The last column reports the size of the devices.
According to the size of a device, which depends on the device technology, some properties are more easily obtained. At nanometric scale, quantum and stochastic activation properties are obtained, but currently associative plasticity has not been demonstrated at such a small scale. For devices of size above 20 nm it is possible to implement plasticity but generally no quantum fluctuation effects are involved.
An exception is provided by the Hf-based RRAMs where both stochastic effects and plasticity have been reported, at the intermediate length scale of about 20 nm. Silicon technology is able to realize long term, short term and associative plasticity only at a circuital level involving several components, while native properties for emulation of plasticity are not reported, so the corresponding elements in Table 2 are left empty. None of the mentioned technologies reproduces pure depression mechanisms. LTD can be achieved as a byproduct of STDP. 
Therefore, if we exclude plasticity of silicon for the above reason, currently none of the materials and the methods reported so far provides the complete set of properties considered above to implement biologically inspired artificial neurons.
It is difficult at the current stage to predict which method will prove to be the most efficient for future hardware implementations of brain-like capabilities, and more research is needed.

%\begin{figure*}[t]%2
%\caption{The Central Server Based Architecture for Data Sharing}\label{fig:integration2}
%\centerline{\epsffile{Figure2.eps}}
%\end{figure*}

%\begin{figure*}[t]%3
%\caption{An Example of Publishing an Existing Document}
%\label{fig:docpublish}
%\centerline{\epsffile{E:/Inderscience/LATEX-FILES/IJMSO/F3.eps}}
%\end{figure*}

\section{Conclusion}
In order to implement biologically inspired properties of real neurons, a number of key ingredients emulating several kinds of plasticity and stochastic effects are required. 
The emulation of neurons and synapses at a more physical level calls for a multi--platform/multi--materials approach to support plasticity. There are suggestions to improve computational capability by including usually unwanted features such as noise, fluctuations, tolerance to large variability, and randomness of nodes and of connections. Many of these properties naturally emerge at the quantum level, which governs the nanometric scale.

\section*{Acknowledgements}
The activity was partially supported by Hokkaido University. The author thanks Tetsuya Asai for useful discussions and the anonymous Reviewer for the proactive revision of the first version of the manuscript and for suggesting to add Table 1.

%%%%%%%%%%%%%%%%%%%%%%%%%%%%%%%%%%%%%%%%%%%%%%%%%%%%%%%%%%%%%%%%%%%%%%%%%%%%%%%%%%%%%

%\bibliography{ijmso}

\begin{thebibliography}{10}

\bibitem{Izhikevitch04}
Izhikevich, E. M. (2004) 'Which model to use for cortical spiking neurons?', \textit{IEEE transactions on neural networks}, Vol. 15, No. 5, pp.1063-1070.

\bibitem{Lizeth12}
Lizeth G.-C., Asai T., and Motomura M. (2012) 'Impact of Noise on Spike Transmission through Serially Connected Electrical FitzHugh-Nagumo Circuits with Subthreshold and Suprathreshold Interconductances', \textit{Journal of Signal Processing}, Vol. 16, No. 6, pp.503--507.

\bibitem{ITRS}
ITRS, http://www.itrs.org/

\bibitem{Indiveri11}
Indiveri, G., et al. (2011), 'Neuromorphic silicon neuron circuits', \textit{Frontiers in neuroscience}, Vol. 5.

\bibitem{Kinjo05}
Kinjo, M., Sato, S., Nakamiya, Y., and Nakajima, K. (2005), 'Neuromorphic quantum computation with energy dissipation', \textit{Physical Review A}, Vol. 72, No. 5, pp.052328.

\bibitem{Maeda00}
Maeda, Y., and Makino, H. (2000) 'A pulse-type hardware neuron model with beating, bursting excitation and plateau potential', \textit{BioSystems}, Vol. 58, No. 1, pp.93--100.

\bibitem{Sahu13}
Sahu, S., Ghosh, S., Hirata, K., Fujita, D., and Bandyopadhyay, A. (2013), 'Multi-level memory-switching properties of a single brain microtubule', \textit{Applied Physics Letters}, Vol. 102, No. 12, pp.123701.

\bibitem{Nori13}
Lambert, N., Chen, Y. N., Cheng, Y. C., Li, C. M., Chen, G. Y., and Nori, F. (2013), 'Quantum biology', \textit{Nature Physics}, Vol. 9, No. 1, pp.10--18.

\bibitem{Escobar07} 
Escobar M.L., Derrick B. (2007) `Long-Term Potentiation and Depression as Putative Mechanisms for Memory Formation', {\it In Bermudez-Rattoni F. Neural plasticity and memory: from genes to brain imaging}, Boca Raton: CRC Press. ISBN 0-8493-9070-2.

\bibitem{LTP}
Morris, R. G. M., Anderson, E., Lynch, G. S., and Baudryl, M. (1986) 'Selective impairment of learning and blockade of long-term potentiation by an N-methyl-D-aspartate' \textit{Nature} Vol. 319, pp. 774--776.

\bibitem{LTD4}
Manahan-Vaughan, D. (2005) 'Hippocampal long-term depression as a declarative memory mechanism', in Synaptic Plasticity and Transsynaptic Signaling (pp. 305-319). Springer US.

\bibitem{LTD3}
Kemp, A., and Manahan-Vaughan, D. (2007) 'Hippocampal long-term depression: master or minion in declarative memory processes?' \textit{Trends in Neurosciences} Vol. 30, No. 3, pp. 111--118.

\bibitem{Purves}
Purves, D. (2008). `Neuroscience (4th ed.)'. Sunderland, Mass: Sinauer. pp. 197–200. ISBN 0-87893-697-1.

\bibitem{LTD1}
Nicholls, R. E., Alarcon, J. M., Malleret, G., Carroll, R. C., Grody, M., Vronskaya, S., and Kandel, E. R. (2008) 'Transgenic mice lacking NMDAR-dependent LTD exhibit deficits in behavioral flexibility', \textit{Neuron}, Vol. 58, No. 1, pp. 104--117.

\bibitem{LTD2}
Malleret, G., et al. (2010) 'Bidirectional regulation of hippocampal long-term synaptic plasticity and its influence on opposing forms of memory', \textit{The Journal of Neuroscience} Vol. 30, No. 10, pp. 3813--3825.

\bibitem{STF1}
Mongillo, G., Barak, O., and Tsodyks, M. (2008) 'Synaptic theory of working memory', \textit{Science} Vol. 319, pp. 1543--1546.

\bibitem{STF2}
Buonomano, D. V., Maass, W. (2009) 'State-dependent computations: spatiotemporal processing in cortical networks', \textit{Nature Reviews Neuroscience} Vol. 10, No. 2, pp. 113--125.

\bibitem{STD}
Goldman, M. S., Maldonado, P., and Abbott, L. F. (2002) 'Redundancy reduction and sustained firing with stochastic depressing synapses', \textit{The Journal of Neuroscience} Vol. 22, No. 2, pp. 584--591.

\bibitem{Sjostrom01}
Sj\"ostr\"om, P.J., Turrigiano, G.G., and Nelson, S.B. (2001), 'Rate, timing, and cooperativity jointly determine cortical synaptic plasticity', \textit{Neuron } Vol. 32, pp.1149--1164.

\bibitem{Corcoran}
Corcoran, T. G., Farmer, S. F., and Berthouze, L. (2010) 'Role of STDP and heterogeneity in the emergence of long-range temporal correlations and frequency scaling in networks of LIF neurons', \textit{BMC Neuroscience}, Vol. 11, Suppl. 1, pp. 23.

\bibitem{STDP3}
Guyonneau, R., VanRullen, R., and Thorpe, S.J. (2005) 'Neurons tune to the earliest spikes through STDP', \textit{Neural Computation}, Vol. 17, pp. 859-879. 

\bibitem{STDP1}
Gerstner, W., Ritz, R., and van Hemmen, J. L. (1993) 'Why spikes? Hebbian learning and retrieval of time-resolved excitation patterns', \textit{Biol. Cybern.} Vol. 69, pp. 503--515.

\bibitem{STDP2}
Gerstner, W., Kempter R., van Hemmen J.L., and Wagner H. (1996), 'A neuronal learning rule for sub-millisecond temporal coding', \textit{Nature}, Vol. 386, pp. 76--78. 

\bibitem{ISTDP1}
Conde, V., et al. (2013) 'Reversed timing-dependent associative plasticity in the human brain through interhemispheric interactions', \textit{Journal of Neurophysiology}, Vol. 109, No. 9, pp. 2260--2271.

\bibitem{ISTDP2}
Requarth, T. and Sawtell, N. B. (2011) 'Neural mechanisms for filtering self-generated sensory signals in cerebellum-like circuits', \textit{Current Opinion in Neurobiology} Vol. 21, No. 4, pp. 602--608.

\bibitem{White00}
White J. A., Rubinstein J. T. and Kay A. R. (2000), 'Channel noise in neurons',  \textit{Trends Neuroscience} Vol. 23, pp.131–-137

\bibitem{Tuckwell10}
Tuckwell H. C., Jost J. (2010), 'Weak Noise in Neurons May Powerfully Inhibit the Generation of Repetitive Spiking but Not Its Propagation',  \textit{PLoS Computational Biology} Vol. 6, No. 5 pp.e1000794

\bibitem{Maass}
Maass, W. (2014) 'Noise as a resource for computation and learning in networks of spiking neurons', \textit{Proceedings of the IEEE}, Vol. 102, No. 5, pp. 860--880.

\bibitem{invivo}
Teich, M. C., Heneghan, C., Lowen, S. B., Ozaki, T., and Kaplan, E. (1997) 'Fractal character of the neural spike train in the visual system of the cat', \textit{JOSA A}, Vool. 14, No. 3, pp. 529--546.

\bibitem{Sobie11}
Sobie, C., Babul, A., and de Sousa, R. (2011). 'Neuron dynamics in the presence of 1/f noise', \textsl{Physical Review E}, Vol. 83, No. 5, pp. 051912.


\bibitem{Kuzum11}
Kuzum, D., Jeyasingh, R. G., Lee, B., and Wong, H. S. P. (2011), 'Nanoelectronic programmable synapses based on phase change materials for brain-inspired computing', \textit{Nano letters}, Vol. 12, No. 5, 2179--2186.

\bibitem{Pershin12}
Pershin, Y. V. and Di Ventra, M. (2012) 'Neuromorphic, digital, and quantum computation with memory circuit elements', \textit{Proceedings of the IEEE}, Vol. 100, No. 6, pp. 2071--2080.

\bibitem{Modha}
Merolla, P. A. et al. (2014). 'A million spiking-neuron integrated circuit with a scalable communication network and interface'. \textit{Science}, Vol. 345, pp.668--673.

\bibitem{Mazzeo12}
Mazzeo G., Prati E., Belli M., Leti G., Cocco S., Fanciulli M., Guagliardo F., Ferrari G. (2012), 'Charge dynamics of a single donor coupled to a few electrons quantum dot in silicon' \textit{Applied Physical Letters} Vol. 100, pp.213107.

\bibitem{Prati12}
Prati, E., Hori, M., Guagliardo, F., Ferrari, G., and Shinada, T. (2012), 'Anderson-Mott transition in arrays of a few dopant atoms in a silicon transistor' \textit{Nature Nanotechnology}, Vol. 7, No. 7, pp. 443--447

\bibitem{Prati12b} Prati E. et al. (2012), 'Few Electron Limit of n-type Metal Oxide Semiconductor Single Electron Transistors', \textit{Nanotechnology} Vol. 23, pp.215204.

\bibitem{Prati07} 
Prati, E., Fanciulli, M., Calderoni, A., Ferrari, G., and Sampietro, M. (2007), 'Microwave irradiation effects on random telegraph signal in a MOSFET', \textit{Physics Letters A}, Vol. 370, No. 5, pp.491--493.

\bibitem{Prati08} Prati E., Fanciulli M., Ferrari G., Sampietro M. (2008) 'Giant Random Telegraph Signal Generated by Single Charge Trapping in sub-micron n-MOSFETs', \textit{Journal Applied Physics} Vol. 103, pp.123707.

\bibitem{Oya06}
Oya T., Asai T., Kagaya R., Hirose T., and Amemiya Y. (2006) 'Neuronal synchrony detection on single-electron
neural networks', \textit{Chaos, Solitons and Fractals}, Vol. 27,  pp.887--894.

\bibitem{Ohno11}
Ohno, T., Hasegawa, T., Tsuruoka, T., Terabe, K., Gimzewski, J. K., and Aono, M. (2011) 'Short-term plasticity and long-term potentiation mimicked in single inorganic synapses', \textit{Nature Materials}, Vol. 10, No. 8, pp. 591--595.

%\bibitem[\protect\citeauthoryear{Oya}{2009}]{Oya09}
%Oya, T (2009) 'Noise-Supported Operations of Neuromorphic Single-Electron Circuits', \textit{International Symposium on Intelligent Signal Processing and Communication Systems} (ISPACS 2009) December 7-9, 2009 TP1-C-3

\bibitem{Nayak12}
Nayak, A., Ohno, T., Tsuruoka, T., Terabe, K., Hasegawa, T., Gimzewski, J. K., and Aono, M. (2012), 'Controlling the Synaptic Plasticity of a Cu2S Gap‐Type Atomic Switch', \textit{Advanced Functional Materials}, Vol. 22, No. 17, pp. 3606--3613.

\bibitem{Shirm12}
Schirm, C., Matt, M., Pauly, F., Cuevas, J. C., Nielaba, P., and Scheer, E. (2013), 'A current-driven single-atom memory', \textit{Nature Nanotechnology}, Vol. 8, No. 9, pp.645--648

\bibitem{Wang12}
Wang, Z. Q., Xu, H. Y., Li, X. H., Yu, H., Liu, Y. C., and Zhu, X. J. (2012), 'Synaptic learning and memory functions achieved using oxygen ion migration/diffusion in an amorphous InGaZnO memristor', \textit{Advanced Functional Materials}, Vol. 22, No. 13, pp.2759--2765.

\bibitem{Puglisi12}
Puglisi, F. M., Pavan, P., Padovani, A., Larcher, L., and Bersuker, G. (2012) `Random telegraph signal noise properties of HfOx RRAM in high resistive state', {\it Solid-State Device Research Conference (ESSDERC), 2012 Proceedings of the European}, pp.274--277. Bordeaux, France

\bibitem{Ambrogio13}
Ambrogio S., Balatti S., Nardi F., Facchinetti S. and Ielmini D. (2013) 'Spike-timing dependent plasticity in a
transistor-selected resistive switching memory', \textit{Nanotechnology } Vol. 24, pp.384012 

\bibitem{Ambrogio15}
Ambrogio, S., Balatti, S., McCaffrey, V., Wang, D. C., and Ielmini, D. (2015) 'Noise-Induced Resistance Broadening in Resistive Switching Memory—Part I: Intrinsic Cell Behavior', \textit{IEEE Transactions on Electron Devices}, Vol. 62, No. 11, pp.3805--3811.

\bibitem{Lai10}
Lai, Q., Zhang, L., Li, Z., Stickle, W. F., Williams, R. S., and Chen, Y. (2010), 'Ionic/electronic hybrid materials integrated in a synaptic transistor with signal processing and learning functions', \textit{Advanced Materials}, Vol. 22, No. 22, pp.2448--2453.

%\bibitem[\protect\citeauthoryear{Ohno~et~al.}{2011}]{OhnoBook}
%Piwowar, H., Becich, M., Bilofsky, H. and Crowley, R. (2011) `PLoS
%medicine', {\it Sept, No. 9, Towards a Data Sharing Culture:
%Recommendations for Leadership from Academic Health Centers},
%Vol.~5.

%\bibitem[\protect\citeauthoryear{Chin and Lansing}{2004}]{chin04context} Chin Jr., G. and Lansing,
%C.S. (2008) `Capturing and Supporting Contexts for Scientific Data
%Sharing via the Biological Sciences Collaboratory', {\it CSCW}, ISBN
%1-58113-810-5.

%\bibitem[\protect\citeauthoryear{Myneni and Patel}{2010}]{myneni10col}
%Myneni, S. and  Patel, V.L. (2010) `Organization of biomedical data
%for collaborative scientific research: A research information
%anagement system', {\it International Journal of Information
%Management}, Vol. 30, No. 3, June, pp.256--264.

\end{thebibliography}
%\bibliographystyle{unsrt}
%\bibliographystyle{alpha}

\end{document}